
%
%
\newfam\scrfam
\batchmode\font\tenscr=rsfs10 \errorstopmode
\ifx\tenscr\nullfont
	\message{rsfs script font not available. Replacing with calligraphic.}
\else	\font\sevenscr=rsfs7
	\font\fivescr=rsfs5
	\skewchar\tenscr='177 \skewchar\sevenscr='177 \skewchar\fivescr='177
	\textfont\scrfam=\tenscr \scriptfont\scrfam=\sevenscr
	\scriptscriptfont\scrfam=\fivescr
	\def\scr{\fam\scrfam}
	\def\cal{\scr}
\fi
\newfam\msbfam
\batchmode\font\twelvemsb=msbm10 scaled\magstep1 \errorstopmode
\ifx\twelvemsb\nullfont\def\Bbb{\bf}
	\message{Blackboard bold not available. Replacing with boldface.}
\else	\catcode`\@=11
	\font\tenmsb=msbm10 \font\sevenmsb=msbm7 \font\fivemsb=msbm5
	\textfont\msbfam=\tenmsb
	\scriptfont\msbfam=\sevenmsb \scriptscriptfont\msbfam=\fivemsb
	\def\Bbb{\relax\expandafter\Bbb@}
	\def\Bbb@#1{{\Bbb@@{#1}}}
	\def\Bbb@@#1{\fam\msbfam\relax#1}
	\catcode`\@=\active
\fi
%
%
\font\eightrm=cmr8	\def\xrm{\eightrm}
\font\eightbf=cmbx8	\def\xbf{\eightbf}
\font\eightit=cmti8	\def\xit{\eightit}
\font\eighttt=cmtt8	\def\xtt{\eighttt}
\font\eightcp=cmcsc8

\font\tentt=cmtt10
\font\twelverm=cmr12
\font\twelvecp=cmcsc10 scaled\magstep1

%
%
\headline={\ifnum\pageno=1\hfill\else
{\eightcp Cederwall, Ferretti, Nilsson, Salomonson:
	``Low Energy Dynamics of Monopoles$\ldots$''}
		\dotfill\folio\fi}
\def\makeheadline{\vbox to 0pt{\vss\noindent\the\headline\break
\hbox to\hsize{\hfill}}
	\vskip2\baselineskip}
%
%
\footline={}
\def\makefootline{\baselineskip=1.6cm\line{\the\footline}}
%
%
\newcount\sectioncount
\sectioncount=0
\def\section#1#2{\global\eqcount=0
	\global\advance\sectioncount by 1
	\vskip2\baselineskip\noindent
	\hbox{\twelvecp\the\sectioncount. #2\hfill}\vskip\baselineskip
	\xdef#1{\the\sectioncount}}
\def\appendix#1{\vskip2\baselineskip\noindent
	\hbox{\twelvecp Appendix: #1\hfill}\vskip\baselineskip}
%
%
\newcount\refcount
\refcount=0
\newwrite\refwrite
\def\ref#1#2{\global\advance\refcount by 1
	\xdef#1{\the\refcount}
	\ifnum\the\refcount=1
	\immediate\openout\refwrite=\jobname.refs
	\fi
	\immediate\write\refwrite
		{\item{[\the\refcount]} #2\hfill\par\vskip-2pt}}
\def\refout{\catcode`\@=11
	\xrm\immediate\closeout\refwrite
	\vskip2\baselineskip
	{\noindent\twelvecp References}\hfill\vskip\baselineskip
	\parskip=.875\parskip
	\baselineskip=.8\baselineskip
	\input\jobname.refs
	\parskip=8\parskip \divide\parskip by 7
	\baselineskip=1.25\baselineskip 
	\catcode`\@=\active}
%
%
\newcount\eqcount
\eqcount=0
\def\Eqn#1{\global\advance\eqcount by 1
	\xdef#1{\the\sectioncount.\the\eqcount}
		\eqno(\the\sectioncount.\the\eqcount)}
\def\eqn{\global\advance\eqcount by 1
	\eqno(\the\sectioncount.\the\eqcount)}
%
%
\parskip=3.5pt plus .3pt minus .3pt
\baselineskip=12pt plus .1pt minus .05pt
\lineskip=.5pt plus .05pt minus .05pt
\lineskiplimit=.5pt
\abovedisplayskip=10pt plus 4pt minus 2pt
\belowdisplayskip=\abovedisplayskip
\hsize=15cm
\vsize=20cm
\hoffset=1cm
\voffset=1.4cm
%
%
\def\/{\over}
\def\*{\partial}
\def\a{\alpha}
\def\b{\beta}
\def\d{\delta}
\def\e{\varepsilon}

\def\l{\lambda}
\def\ld{\lambda^\dagger}
\def\p{\psi}

\def\r{\varrho}

\def\w{\omega}
\def\G{\Gamma}

\def\S{\Sigma}
\def\tS{\tilde\Sigma}
\def\Z{{\Bbb Z}}
\def\C{{\Bbb C}}
\def\R{{\Bbb R}}
\def\H{{\Bbb H}}
\def\L{{\cal L}}
\def\M{{\cal M}}
\def\Ham{{\cal H}}
\def\ad{\hbox{ad}\,}
\def\punkt{\,\,.}
\def\komma{\,\,,}
\def\.{.\hskip-1pt }
\def\is{\!=\!}
\def\-{\!-\!}
\def\+{\!+\!}
\def\Int{\int\limits}
\def\bra{\,<\!\!}
\def\ket{\!\!>\,}
\def\ra{\rightarrow}
\def\half{{1\/2}}
\def\cross{\!\times\!}
\def\Int{\int\!d^3x\,}
\def\Tr{\hbox{Tr}\,}
\def\ker{\hbox{ker}\,}
\def\f{{\lower1.5pt\hbox{$\scriptstyle f$}}}
\def\ff{{\lower1pt\hbox{$\scriptstyle f$}}}
\def\Re{\hbox{Re}\,}
%
%
%
%
%
\null\vskip-1cm
\hbox to\hsize{\hfill G\"oteborg-ITP-95-16}
\hbox to\hsize{\hfill\tt hep-th/9508124}
\hbox to\hsize{\hfill August, 1995}

\vskip4cm

\vskip2cm
\centerline{\twelvecp Low Energy Dynamics of Monopoles in N=2 SYM with Matter}
\vskip\parskip
\centerline{\twelvecp}

\vskip1.5cm
\centerline{\twelverm Martin Cederwall, Gabriele Ferretti, Bengt E.W. Nilsson
	and Per Salomonson}

\vskip1cm
\centerline{\it Institute for Theoretical Physics}
\centerline{\it Chalmers University of Technology and G\"oteborg University}
\centerline{\it S-412 96 G\"oteborg, Sweden}

\vskip1.5cm
\noindent \underbar{Abstract:} We derive, for $N\is2$ super-Yang--Mills with
gauge group $SU(2)$ and massless matter, the supersymmetric
quantum mechanical models describing the time evolution of multi-monopole
configurations in the low energy approximation.
This is a first step towards
identifying the solitonic states mapped to fundamental excitations
by duality in the model with four hypermultiplets in the fundamental
representation.
\vskip1cm
\vfill
\catcode`\@=11
\hbox to\hsize{email addresses: \tentt tfemc@fy.chalmers.se\hfill}
\hbox to\hsize{\phantom{email addresses: }\tentt ferretti@fy.chalmers.se\hfill}
\hbox to\hsize{\phantom{email addresses: }\tentt tfebn@fy.chalmers.se\hfill}
\hbox to\hsize{\phantom{email addresses: }\tentt tfeps@fy.chalmers.se\hfill}
\catcode`\@=\active

\eject

\ref\SeibergIV{N.~Seiberg and E.~Witten,
	\xit Nucl.Phys. \xbf B426 \xrm(1994) 19;
	\xrm erratum: ibid. \xbf B430 \xrm (1994) 485 ({\xtt hep-th/9407087}).}
\ref\SeibergIII{N.~Seiberg and E.~Witten, \xit Nucl.Phys. \xbf B431
	\xrm (1994) 484 ({\xtt hep-th/9408099}).}
\ref\Vafa{C.~Vafa and E.~Witten, \xit Nucl.Phys. \xbf B431 \xrm (1994) 3
	({\xtt hep-th/9408074}).}
\ref\IntriligatorI{K.~Intriligator and N.~Seiberg \xit Nucl.Phys. \xbf B431
	\xrm (1994) 551 ({\xtt hep-th/9408155}).}
\ref\Montonen{C.~Montonen and D.~Olive,
	\xit Phys.Lett. \xbf 72B \xrm (1977) 117.}
\ref\Font{A. Font, L.E. Iba\~nez, D. L\"ust and F. Quevedo,
	\xit Phys.Lett. \xbf 249B \xrm (1990) 35.}
\ref\Mandelstam{S.~Mandelstam, \xit Nucl.Phys. \xbf B213 \xrm (1983) 149.}
\ref\Brink{L.~Brink, O.~Lindgren and B.E.W.~Nilsson,
	\xit Phys.Lett. \xbf 123B \xrm (1983) 323.}
\ref\Sohnius{M.~Sohnius and P.~West, \xit Phys.Lett. \xbf 100B \xrm (1981) 45.}
\ref\Howe{P.S.~Howe, K.S.~Stelle and P.C.~West, \xit Phys.Lett. \xbf 124B
	\xrm (1983) 55.}
\ref\HoweII{P.S.~Howe, K.S.~Stelle and P.K.~Townsend,
	\xit Nucl.Phys. \xbf B214 \xrm (1983) 519.}
\ref\Piguet{O.~Piguet and K.~Sibold,
	\xit Helv.Phys.Acta \xbf 63 \xrm (1990) 71.}
\ref\SeibergV{N.~Seiberg \xit Phys.Lett. \xbf 206B \xrm (1988) 75.}
\ref\Osborn{H.~Osborn, \xit Phys.Lett. \xbf 83B \xrm (1979) 321.}
\ref\Bogomolny{E.B.~Bogomolny, \xit Sov.J.Nucl.Phys. \xbf 24 \xrm (1976) 449.}
\ref\Prasad{M.K.~Prasad and C.M.~Sommerfield, \xit Phys.Rev.Lett. \xbf 35
	\xrm (1975) 760.}
\ref\MantonI{N.S.~Manton \xit Phys.Lett. \xbf 110B \xrm (1982) 54.}
\ref\Christ{N.H.~Christ and T.D.~Lee,
	\xit Phys.Rev. \xbf D22 \xrm (1980) 939.}
\ref\Rajaraman{R.~Rajaraman, \xit ``Solitons and Instantons'',
	\xrm North-Holland (1982).}
\ref\Wess{J.~Wess and J.~Bagger, \xit ``Supersymmetry and Supergravity'',
	\xrm Princeton Univ. Press (1983).}
\ref\Kugo{T.~Kugo and P.~Townsend, \xit Nucl.Phys. \xbf B221 \xrm (1983) 357.}
\ref\Sudbery{A.~Sudbery, \xit J.Phys. \xbf A17 \xrm (1984) 939.}
\ref\AtiyahII{M.F.~Atiyah and I.M.~Singer,
	\xit Proc.Nat.Acad.Sci.USA \xbf 81 \xrm (1984) 2597.}
\ref\MantonII{N.S.~Manton and B.J.~Schroers, \xit Ann.Phys. \xbf 225
	\xrm (1993) 290.}
\ref\Blum{J.~Blum, \xit Phys.Lett. \xbf 333B \xrm (1994) 92
	({\xtt hep-th/9401133}).}
\ref\GauntlettII{J.P.~Gauntlett, \xit Nucl.Phys. \xbf B411 \xrm (1994) 443
	({\xtt hep-th/9305068}).}
\ref\SenI{A.~Sen, \xit Phys.Lett. \xbf B329 \xrm (1994) 217
	({\xtt hep-th/9402032}).}
\ref\Parkes{A.~Parkes and P.C.~West,
	\xit Phys.Lett. \xbf 122B \xrm (1983) 365.}
\ref\Adda{A.~D'Adda, R.~Horsley and P.~Di~Vecchia, \xit Phys.Lett.
	\xbf 76B \xrm (1978) 298.}
\ref\AtiyahI{M.F.~Atiyah and N.~Hitchin, \xit ``The Geometry and Dynamics of
	Magnetic Monopoles'',\hfill\break\indent
	 \xrm Princeton University Press, Princeton (NJ) (1988).}
\ref\Harvey{J.A.~Harvey and A.~Strominger, \xit Commun.Math.Phys. \xbf
	151 \xrm (1993) 221 ({\xtt hep-th/9108020}).}
\ref\Jackiw{R.~Jackiw and C.~Rebbi,
	\xit Phys.Rev. \xbf D13 \xrm (1976) 3398.}
\ref\Callias{C.~Callias, \xit Commun.Math.Phys. \xbf 62 \xrm (1978) 213.}
\ref\WittenV{E.~Witten, \xit Phys.Rev. \xbf D16 \xrm (1977) 2991.}
\ref\WittenI{E.~Witten, \xit Nucl.Phys. \xbf B202 \xrm (1982) 253.}
\ref\Vecchia{P.~Di~Vecchia and S.~Ferrara,
	\xit Nucl.Phys. \xbf B130 \xrm (1977) 93.}
\ref\Bagger{J.~Bagger, Lectures at Bonn-NATO Adv.
	Study Inst. on Supersymmetry, 1984, \hfill\break\indent
	\xit Bonn Supersym.ASI \xrm (1984) 45.}
\ref\Alvarez{L.~Alvarez-Gaum\'e,
	\xit Commun.Math.Phys. \xbf 90 \xrm (1983) 161.}
\ref\GauntlettI{J.P.~Gauntlett, \xit Nucl.Phys. \xbf B400 \xrm (1993) 103
	({\xtt hep-th/9205008}).}
\ref\Sehti{S.~Sethi, M.~Stern and E.~Zaslow, {\xtt hep-th/9508117}.}
\ref\GauntlettIII{J.P.~Gauntlett and J.A.~Harvey, {\xtt hep-th/9508156}.}

\section\introduction{Introduction}

The holomorphicity appearing in supersymmetric gauge theories provides a
unique opportunity to understand their non-perturbative structures. This has
been amply demonstrated in a number of recent papers where
asymptotically free (or ultraviolet finite) Yang--Mills
theories with $N\is1,2$ or $4$ supersymmetries coupled to a variety of
matter multiplets have been shown to possess a number of interesting
exact properties [\SeibergIV
-\IntriligatorI].

One of the most intriguing features that can now be investigated is the
duality between strong and weak coupling. The possibility in field theory
of such a property was first discussed by Montonen and Olive [\Montonen]
and in the context
of string theory by Font et al. [\Font].

For $N\is1,2$, duality relates different low energy descriptions
of the same theory
at different points in the quantum moduli space. While in these cases duality
is not a symmetry of the theory, the situation changes drastically
if we instead consider
theories with vanishing $\beta$-function, e.g\.\ $N\is4$ super-Yang--Mills.
This theory
is known to be ultraviolet finite at the perturbative
[\Mandelstam
-\Piguet] as
well as non-perturbative [\SeibergV] level. Here one can perform a duality
transformation from weak to strong coupling and then tune to weak coupling
thereby making a comparison possible. This
theory was conjectured to be selfdual in ref\.\ [\Osborn].

As emphasized recently by Seiberg and Witten [\SeibergIII], there is
another candidate, perhaps more interesting, for
such a selfdual theory, namely $SU(2)$ $N\is2$ super-Yang--Mills coupled to
four hypermultiplets in the fundamental representation of the gauge group.
Such $N\is2$ theories are perturbatively finite provided the number of
hypermultiplets $N_f$ and the gauge group $SU(N_c)$ are related by
$N_f\is2N_c$. It was argued in [\SeibergIII] that this is
most likely true also non-perturbatively. The purpose of the present paper
is to derive the quantum mechanical models governing the dynamics
of the zero-modes obtained from the $N_c\is2$ theory with hypermultiplets
in the background of a BPS (multi-)monopole [\Bogomolny,\Prasad].

The required steps needed to obtain the lagrangian for the zero-modes
are well-known and can be found in many places in the literature;
see e.g\.\ [\MantonI,\Christ,\Rajaraman]. Here we will use a somewhat
unusual formulation of the $N\is2$ theory based on quaternions.
There are several reasons why quaternions are convenient in this context.
One is that the lagrangian in $D\is(1\!+\!3)$ dimensions is most easily
derived from the $N\is1$, $D\is(1\!+\!5)$ analogue, and that this theory
has a nice formulation in terms of quaternions. Just as the
isomorphism between $Spin(1,3)$ and $Sl(2,\C)$ makes it
possible  to introduce the extremely useful van der Wearden notation (see
e.g\.\ the book by Wess and Bagger [\Wess]), in $D\is(1\!+\!5)$ the fact
that $Spin(1,5)$
and $Sl(2,\H\,)$ are isomorphic [\Kugo,\Sudbery] leads to
a  two-component formulation quite similar to the one in $D\is(1\!+\!3)$ but
where the ``Pauli matrices'' are now five in number due to their
entries being quaternionic. This is explained in section 2. A second reason
is that many properties of monopoles become particularly apparent
if discussed in terms of three dimensional space accompanied by a fourth
euclidean dimension as e.g\.\ in the context of
the Atiyah-Singer universal bundle [\AtiyahII].
This extra dimension is just a fourth space dimension of the
$(1\!+\!5)$-dimensional theory we start from, and as we will see
there is actually
no need for giving the theory in $(1\!+\!3)$ dimensions at all.

Quaternions are in fact already widely used in connection with monopoles
but for a reason slightly different from ours. When seeking explicit
functional forms of the zero modes and the monopole solutions themselves
it has in the past turned out to be very fruitful to conduct the search
in the language of quaternions [\MantonII].

In section 3 we present the quaternionic formulation in
$D\is(1\!+\!5)$ of the relevant
super-Yang--Mills theory coupled to hypermultiplets, while section 4 is devoted
to the derivation of the zero-mode formulas needed to integrate the
lagrangian over three-space. We give e.g\.\ exact expressions for the curvature
of moduli space as well as the curvature of the relevant index bundle
that appears due to the presence of new fermionic zero-modes coming
from the hypermultiplets.

The supersymmetry of the quantum mechanical model for the zero-modes
is discussed in detail in section 5. The lagrangian found here
coincides with the one obtained in section 6 by integrating the
field theory lagrangian of section 3 over space using the monopole
formulas in section 4. Our results are consistent with the ones stated
by Blum [\Blum] for the $N\is4$ theory and extends the ones of Gauntlett
[\GauntlettII] for $N\is2$
by including hypermultiplets.
We plan to continue these investigations in a future publication.
In section 7, we outline the way to proceed in order to find the
necessary forms
on moduli space that would vindicate the conjecture about selfduality of
this theory along the lines already used by Sen [\SenI] for the $N\is4$ theory.

\section\quaternions{Quaternionic notation}

Even though the concept of quaternions is well known among physicists,
not many are accustomed to actually using them in calculations.
This preliminary section may be skipped over by the reader already
familiar with the use of quaternions.
The reason we choose a quaternionic notation throughout this paper is
twofold: firstly, it provides a natural and compact formulation of
supersymmetric Yang--Mills theory in six dimensions, the easiest way
to $N\is2$ in four dimensions, secondly, they arise naturally in
the context of monopoles and their moduli. Using a quaternionic
notation in $D\is(1\!+\!5)$ enables us to go directly to a most elegant
version of monopole theory, formulated on a euclidean four-dimensional
space, without explicitly considering $(1\!+\!3)$-dimensional Minkowski space.

A quaternion $h\in\H$ is a collection of four real numbers $h_a$,
$a=1,\ldots,4$, arranged as $h=h_ae_a$, where $e_4\is1$,
$e_ie_j\is-\d_{ij}+\e_{ijk}e_k$, $i,j\is1,2,3$. Under quaternionic conjugation,
denoted by $^*$, $h\ra h^*\is h_4-h_ie_i$. Conjugation is an anti-involution,
i.e\.\ $(hh')^*\is h'^*h^*$. We denote the real part by
$\Re h\is h_4\is\half(h+h^*)$. The norm of a quaternion is given by
$|h|^2\is hh^*\is h_ah_a$. The unit quaternions
form the group $SU(2)$.

In four euclidean dimensions, a quaternion can represent a vector or any of the
two spinor chiralities. The transformation rules for vectors $v$ and
spinors $s$ and $c$ under the rotation group $SU(2)\cross SU(2)$ are:
$$
\eqalign{
&v\ra hvg^*\komma\cr
&s\ra hs\komma\cr
&c\ra gc\komma\cr}\eqn
$$
where $h$ and $g$ are unit norm quaternions.
The group of three-dimensional rotations is the diagonal subgroup.
These transformations have been chosen in order to satisfy a triality,
formally identical to the one for the three octonionic representations
of $Spin(8)$, i.e\.\ "$v^*s\is c$" and cyclic under $v\ra s^*\ra c\ra v$.
A scalar is formed as $\Re(v'v^*)$ or similarly for the spinors
(i.e\.\ the norm is covariant).
Note that there is room for yet another $SU(2)$ transforming $s$ and $c$
from the right. Such "extra" symmetries will appear as
global symmetries in the field theory.

Gamma matrices in four euclidean dimensions are
$$
\G_\mu=\left[\matrix{0&e_\mu\cr e^*_\mu&0\cr}\right]\komma\eqn
$$
so the Weyl equations for the two spinor chiralities may be written
$$
\eqalign{
&D^*\!s=0\komma\cr
&Dc=0\komma\cr}\eqn
$$
where $D\is e_\mu D_\mu$.
We also note that quaternions $a\is vv'^*$ and $b\is v^*v'$
(and also corresponding expressions with spinors) have
well-defined transformation properties. Their real parts are scalars
and their imaginary parts contain a selfdual and an anti-selfdual
anti-symmetric tensor, respectively.
An example to be used is the covariant equation
$D^*\!v\is0$, that in components reads (antisymmetrization always has weight
one)
$$
\eqalign{
&D_\mu v_\mu=0\komma\cr
&D_{[\mu}v_{\nu]}-\half\e_{\mu\nu\kappa\l}D_\kappa v_\l=0\punkt\cr}
	\Eqn\antiselfdual
$$

Concerning $(1\!+\!5)$-dimensional Minkowski space, there is a two-component
spinor notation relying on the isomorphism
$Spin(1,5)\!\approx\!Sl(2;\H\,)$ [\Kugo,\Sudbery].
The formalism is quite analogous to the $Sl(2;\C)$ spinor notation in
$D\is(1\!+\!3)$. A difference is that there is no ``$\e_{\a\b}$'', so spinor
indices can not be lowered and raised. Therefore, there are two inequivalent
spinor chiralities.
In terms of the euclidean spinors, a six-dimensional spinor
(in any of the two chiralities) is
$$
S=\left[\matrix{s\cr c\cr}\right]\punkt\eqn
$$
The six-dimensional sigma matrices are ($M\is0,\ldots,5$)
$$
\matrix{&\S^0=\left[\matrix{1&0\cr0&1\cr}\right]\komma\hfill
	&\S^\mu=\left[\matrix{0&e_\mu\cr e^*_\mu&0\cr}\right]\komma\hfill
	&\S^5=\left[\matrix{1&0\cr0&-1\cr}\right]\komma\hfill\cr
		&&&\phantom{x}\cr
	&\tS^0=\left[\matrix{1&0\cr0&1\cr}\right]\komma\hfill
	&\tS^\mu=-\left[\matrix{0&e_\mu\cr e^*_\mu&0\cr}\right]\komma\hfill
	&\tS^5=-\left[\matrix{1&0\cr0&-1\cr}\right]\komma\hfill\cr}
	\Eqn\sigmamatrices
$$
and Lorentz transformations act by left multiplication by quaternionic
matrices $\S^{MN}\is\S^{[M}\tS^{N]}$ or
$\tS^{MN}\is\tS^{[M}\S^{N]}$. Again there is room for an extra
commuting $SU(2)$ acting from the right.
If one instead of reducing to four euclidean dimensions goes to
$(1\!+\!3)$-dimensional Minkowski space, the spinors are acted on by left
multiplication by complex matrices generating $Sl(2;\C)$. Thereby
the quaternions are split into $\C\oplus\C$, so that $S$ contains two
four-dimensional Weyl spinors.

\section\nistwo{N=2 super-Yang--Mills with matter multiplets}

The action for $D\is(1\!+\!5)$ super-Yang--Mills with massless matter is
$$
\eqalign{\L=-{1\/4}F_{MN}&F^{MN}+\half\Re(\ld\S^MD_M\l)\cr-
		&\half D_Mq^*_\ff D^Mq_\f
	+\half\Re(\p^\dagger_f\tS^MD_M\p_\f)+\Re(\p^\dagger_f\l q^*_\ff)
	+{1\/8}(q^*_\ff\cross q_\f)^2\punkt\cr}
	\Eqn\fieldtheoryaction
$$
For the sake of compactness of notation, we suppress representation
indices for the gauge
group, which throughout this paper is $SU(2)$. Here, the gauge potential
$A_M$ and the fermion $\l$ form the vector multiplet, and transform
in the adjoint representation of the gauge group. $\l$ is a two-component
quaternionic spinor of the type discussed above. The fields $q$ and $\p$
form the matter hypermultiplets: the index $f$ labels the multiplets, which
each can transform in any representation of the gauge group.
In practice, the only candidates are the adjoint and fundamental,
if asymptotic freedom is to be retained. In the case of adjoint multiplets,
the maximum number is then one, which gives the $N\is2$ theory
in $D\is(1\!+\!5)$,
which can be dimensionally reduced to $N\is4$ in $D\is(1\!+\!3)$. In the case
of fundamental matter, $f\is1,\ldots,N_f$, where $N_f\!\leq\!4$. The
upper limits give finite theories
[\Mandelstam
-\SeibergV,\SeibergIII,\Parkes].
The matter bosons $q$ are Lorentz scalar quaternions, and the fermions $\p$ are
again two-component quaternionic spinors, of the opposite (six-dimensional)
chirality compared to $\l$. The cross product in the last term denotes
the formation of an element in the adjoint.

The supersymmetry transformations are:
$$
\matrix{\d A_M=\Re(\e^\dagger\S_M\l)\komma\hfill
		&\d q_\f=\p^\dagger_f\e\komma\hfill\cr
	\lower4pt\hbox{$\d\l=-\half F_{MN}\tS^{MN}\e
		+\half\e(q^*_\ff\cross q_\f)\komma$}\hfill
	&\lower4pt\hbox{$\d\p_\f=\S^M\e D_Mq^*_\ff\punkt$}\hfill\cr}
	\Eqn\supersymmetrytransformations
$$
The parameter $\e$ is in the same spinor representation as $\l$ and transforms
the same way under global symmetries (see below). The supersymmetry algebra
closes modulo equations of motion for the fermions, since no auxiliary
fields are present in (\fieldtheoryaction).

There is a global $SU(2)\cross SU(2)$ symmetry of this action, transforming the
fields as
$$
\matrix{A_M\ra A_M\komma\hfill	&q\ra g^*qh\komma\hfill\cr
	\lower4pt\hbox{$\l\ra\l h\komma$}\hfill
	&\lower4pt\hbox{$\p\ra\p g\punkt$}\hfill\cr}\eqn
$$
The simplest way of dimensionally reducing to the $N\is2$ action in
$D\is(1\!+\!3)$ [\Adda] is to keep
the spinors quaternionic. Since Lorentz transformations only act with left
multiplication by complex matrices, they effectively split into pairs of
$Sl(2;\C)$ spinors. Also, the vectors split into vectors and scalars.
For our purposes, this is not convenient to do explicitly, since
even though monopole physics takes place in four-dimensional Minkowski space,
the best mathematical formulation makes use of a four-dimensional
euclidean space, where a Higgs field is taken as the fourth component
of the vector potential. This is achieved directly from the six-dimensional
action by singling out the four euclidean dimensions labeled by $\mu$.

\section\moduli{Monopole moduli}

Our aim is to identify the bosonic and fermionic moduli (zero-modes) around
a multi-monopole solution, and write down an action for slow (adiabatic)
motion of these moduli.
A (static) multi-monopole solution is a gauge field configuration
$A_\mu(x)$ where
$\mu\is1,2,3,4$, independent of $x^4$, whose field strength
is selfdual. The components $A_0$ and $A_5$
as well as all other fields vanish. The Higgs field $A_4$ takes a non-zero
value at spatial infinity, breaking the gauge $SU(2)$ to $U(1)$. Half of
the supersymmetry is broken by a monopole solution (see below).
We will not go into great detail on monopole theory
(see [\AtiyahI] and references therein), except for some
information about their moduli spaces needed in this paper. The total moduli
space decomposes into the disjoint union of manifolds of solutions with
different (integer) magnetic charge $k$. All statements about dimensionalities
of moduli spaces and spaces of solutions are made for positive $k$, and are
the same for $k$ negative. The notation is mostly
a quaternionic version of the one in ref\.\ [\Harvey].
The reader should not confuse our use of quaternions
with that in [\MantonII], where the $SU(2)$ gauge generators act with
right quaternionic multiplication in the fundamental representation.

A tangent vector to the moduli space must satisfy
$$
\eqalign{
&D_\mu \d_m A_\mu=0\komma\cr
&D_{[\mu}\d_m A_{\nu]}-\half\e_{\mu\nu\kappa\l}D_\kappa \d_m A_\l=0\cr}
	\Eqn\tangenteq
$$
(the second equation stating that $F_{\mu\nu}$ remains selfdual, the
first one making the tangent vectors orthogonal to gauge modes in the
metric stated below),
which can be written in quaternionic notation as (see eq\.\ (\antiselfdual))
$$
D^*\d_m A=0\komma\eqn
$$
where $\d_m A\is \d_m A_\mu e_\mu$.
We immediately recognize that if $\d_m A$ is a tangent vector, then so is
$\d_m Ah$ for any quaternion $h$. This quaternionic structure of the zero-mode
equation (acting on the scalar and anti-selfdual components of (\tangenteq))
is translated into a quaternionic structure acting on the tangent bundle
of moduli space. The moduli space is hyper-K\"ahler, with the three complex
structures induced as
$$
J^{(i)\phantom{m}n}_{\phantom{(i)}\raise4pt\hbox{$\scriptstyle m$}}
	\d_n A=\d_m Ae_i\punkt\Eqn\complexstructures
$$
It is well known that the moduli space $\M_k$ of $k$-monopole solutions
(modulo gauge transformations) is a $4k$-dimensional hyper-K\"ahler
manifold.

A preferred metric is
$$
g_{mn}=\Int\Tr\d_m A_\mu\d_n A_\mu=\Int\Tr\Re(\d_m A^*\d_n A)
	\equiv\bra\d_m A,\d_n A\ket\punkt
\Eqn\modulimetric
$$
This metric is natural because it derives from the kinetic term of the
gauge field, in the sense that slow motion on moduli space is geodesic
with respect to it (see sections 5 and 6).
One can in fact collect the metric and the complex structures in
$$
\bra\d_m A,\d_n Ah^*\ket=h_4g_{mn}
+h_iJ^{(i)\phantom{m}p}_{\phantom{(i)}\raise4pt\hbox{$\scriptstyle m$}}g_{pn}
=h_a{J^{(a)}}_{mn}\punkt\eqn
$$
where $h$ is a quaternion.

The tangent vectors can be written as
$$
\d_m A=\*_m A-D\e_m\komma\eqn
$$
where the gauge parameters $\e_m$ are chosen so that
the first equation in (\tangenteq) is satisfied.
The operator $s_m\is\*_m+\ad\e_m$ together with $D_\mu$ forms a covariant
derivative on the
``universal bundle'' $\M_k\cross\R^4$ [\AtiyahII].
It fulfills
$$
[s_m,D]=\d_m A\eqn
$$
and Jacobi identities show among other things that
$$
\phi_{mn}\equiv[s_m,s_n]=2(D^*\!D)^{-1}\Re(\d_mA^*\cdot\d_nA)\komma\eqn
$$
where the dot indicates the adjoint action in the gauge $SU(2)$.

The Christoffel connection of the metric (\modulimetric) is
$$
{\G^m}_{np}=g^{mq}\Int\Tr\d_q A_\mu s_n\d_p A_\mu
	=g^{mq}\Int\Tr\Re(\d_q A^*s_n\d_p A)
	=g^{mq}\!\bra\d_q A,s_n\d_p A\ket\punkt
	\eqn
$$
Using that zero-modes and higher modes together form a complete set,
and that zero-modes are orthogonal to higher modes, the validity of this
formula can be verified, as well as the fact that the complex structures
are covariantly constant. When calculating the curvature tensor, one obtains
$$
R_{mnpq}=\bra\d_p A,\phi_{mn}\d_q A\ket+\bra s_m\d_p A,\Pi_+s_n\d_q A\ket
				-\bra s_n\d_p A,\Pi_+s_m\d_q A\ket\komma
	\Eqn\curvature
$$
where $\Pi_+$ is the projection operator on higher modes.
There is in general no reason for
the ``extra'' terms apart from the curvature
of the universal bundle sandwiched with zero-modes to vanish,
a fact that does not seem to be universally recognized in the earlier physics
literature.
In fact, we can use the expression for the projector on higher modes,
$$
\Pi_+\is D(D^*\!D)^{-1}\!D^*\komma\Eqn\projector
$$
to obtain a more explicit expression in terms of the geometrical objects
defined above. A crucial ingredient in this calculation is the ``lifting''
of the complex structures from $\R^4$ to $\M_k$ of (\complexstructures).
The result is
$$
R_{mnpq}=\bra\d_p A,\phi_{mn}\d_q A\ket
	-4{P_{+pq}}^{rs}\!\bra\d_mA,\phi_{nr}\d_sA\ket\komma\Eqn\finalR
$$
where
$$
{P_{+pq}}^{rs}={1\/4}{{J^{(a)}}_{[p}}^{\lower2pt\hbox{$\scriptstyle r$}}
	{{J^{(a)}}_{q]}}^{\lower2pt\hbox{$\scriptstyle s$}}\eqn
$$
($J^{(4)}$ is the identity) is the projection operator on the part of
an antisymmetric tensor that commutes with the complex structures,
i.e\.\ the $Sp(k)$ part. The first term in (\finalR) automatically
lies in this subspace. This expression ensures that the curvature on $\M_k$ has
the correct holonomy and cyclic symmetry.
The projection is a suitable generalization of the
selfduality concept in four dimensions.
As we will see, an analogous term arises in the effective
action for the moduli of $N\is2$ super-Yang--Mills with matter.

When we consider fermions coupled to a multi-monopole
background [\Jackiw,\MantonII,\Callias], the Dirac
equation in four euclidean dimensions reads
$$
\left[\matrix{0&D\cr D^*&0\cr}\right]\left[\matrix{s\cr c\cr}\right]=0\punkt
	\Eqn\diraceqn
$$
Using the selfduality of the field-strength, one sees that $D^*\!D$ is the
Laplace operator, while $DD^*$ also contains the field-strength
as its imaginary part. Therefore,
$\ker D\subseteq\ker D^*\!D=\{0\}$, and the zero-modes must be in the $s$
representation. The index theorem [\Callias] determines their number.
The euclidean Dirac equation (\diraceqn) is the linearized time-independent
equation of motion for the spinors $\l$ and $\p$ of the (dimensionally reduced)
action (\fieldtheoryaction), i.e\.\ it determines their zero-modes.
At this level, the distinction between the six-dimensional chiralities
is lost. However, the two spinor chiralities couple to $A_0$ and $A_5$
with different relative signs (see eq\.\ (\sigmamatrices)), and this prevents
us from making a consistent truncation of the equations of motion
to all orders for the time-independent solutions, as is done in [\GauntlettII]
for pure $N\is2$. This difference will manifest itself in the relative
signs of the fermion bilinears of eq\.\ (6.1).

Considering the expression
$$
\tS_{MN}F^{MN}=-\G_{\mu\nu}F_{\mu\nu}
	=-\left[\matrix{e_\mu e^*_\nu F_{\mu\nu}&0\cr0&0\cr}\right]
	\eqn
$$
occurring in the supersymmetry transformations (\supersymmetrytransformations),
one concludes that the unbroken supersymmetry has transformation parameter
$\e\is\left[\matrix{0\cr\e_c\cr}\right]$.

In the case of adjoint spinors, the Weyl equation is formally identical
to the equation for the bosonic moduli, and it is satisfied by
$$
s\is\d_m A\l^m\eqn
$$
for any set of quaternionic spinors $\l^m$. As for the
bosonic moduli, these are related by the complex structures, so that
$\l^m$ can be chosen real (in the case of real fermions). They equal the
bosonic moduli in number, i.e\.\ the mode functions span a $4k$-dimensional
vector space over $\R$. For quantization, it will be essential to use
complex fermions, and view it as a $2k$-dimensional vector space
over $\C$.
This is the conventional statement that there are $2k$ (complex) zero-modes
in the adjoint representation [\Jackiw,\AtiyahI,\Callias].
Part of the adjoint zero-modes in
$N\is2$ super-Yang--Mills (in the $k\is1$ sector all of them) are Goldstone
fermions, generated by the broken supersymmetry.

For spinors in any representation, the zero-modes, properly normalized, form an
orthonormal bundle over $\M_k$. Fundamental fermions will of course be
of special interest -- then the number of zero-modes is $2k$ real,
i.e\.\ $k$ complex [\Jackiw,\MantonII,\Callias].
We set
$$
s\is\r_\a\p^\a\komma\eqn
$$
where $\r_\a$ are the quaternionic mode functions.
The natural connection on this index bundle is
$$
\w_{m\a\b}=\bra\r_\a,s_m\r_\b\ket\komma\eqn
$$
with $\e_m$ now acting in the appropriate representation of the gauge group,
and its curvature can be calculated completely analogously to above:
$$
F_{mn\a\b}=\bra\r_a,[s_m,s_n]\r_\b\ket
	+\bra s_m\r_\a,\Pi_+s_n\r_\b\ket
	-\bra s_n\r_\a,\Pi_+s_m\r_\b\ket\punkt\Eqn\indexcurvature
$$
where again the covariant derivative contained in the projection operator
(\projector) acts in the appropriate representation.

\section\ssqm{Supersymmetric quantum mechanics}

Semiclassical adiabatic motion of the monopole moduli for supersymmetric
Yang--Mills is governed by
supersymmetric quantum mechanics, i.e\.\ a one-dimensional supersymmetric
sigma model [\WittenV
-\Alvarez],
with moduli space as target space.
On any riemannian manifold, the lagrangian
$$
\L=\half g_{mn}(X)\dot X^m\dot X^n+\half g_{mn}(X)\l^mD_t\l^n\eqn
$$
is supersymmetric. The fermions $\l$ are real, and the covariant derivative
is defined as $D_t\l^m\is\dot\l^m+{\G^m}_{np}(X)\dot X^n\l^p$.
The supersymmetry
is of the type $N\is\half$, i.e\.\ there is only one, real, supersymmetry
generator. The structure of this model, as well as its generalizations
to higher $N$ and number of fermions, is most easily understood in a
canonical framework. Due to the second class constraints between the
fermions and their momenta, the Dirac procedure yields nonvanishing
brackets $\{P_m,\l^n\}$ and $\{P_m,P_n\}$. The most natural phase space
variable to use instead of the canonical momentum is the velocity
$V_m=g_{mn}\dot X^n$. One obtains Dirac brackets
$$
\eqalign{
&\{V_m,X^n\}=-\d_m^n\komma\hfill\cr
&\{V_m,V_n\}=\half R_{mnpq}\l^p\l^q\komma\hfill\cr
&\{V_m,\l^n\}={\G^n}_{mp}\l^p\komma\hfill\cr
&\{\l^m,\l^n\}=g^{mn}\punkt\hfill\cr}\Eqn\brackets
$$
This structure is quite unique -- the Jacobi identities demand that
$g$ is covariantly constant, that $R$ has the usual expression in terms
of $\G$ and that the Bianchi identities for $R$ are fulfilled.
One may as well use an orthonormal basis
for the fermions. Then the third equation in (\brackets) contains the spin
connection instead of the Christoffel connection.

The $N\is\half$ supersymmetry generator is
$$
Q=\l^mV_m\punkt\Eqn\halfgenerator
$$
It is not too difficult to see that if there are covariantly constant
complex structures, each of these gives another supersymmetry generator
$Q_J\is\l^m{J_m}^nV_n$. In the case of a hyper-K\"ahler manifold, the
number of supersymmetries will be multiplied by four. If the fermions
$\l$ are complex, or equivalently, carry an index $i\is1,2$, the
($N\is 1$) supersymmetry generators are $Q^i\is\l^{im}V_m$ and anticommute
to $\{Q^i,Q^j\}\is2\d^{ij}\Ham$, where
$$
\Ham=\half g^{mn}V_mV_n+{1\/8}R_{mnpq}\l^{im}\l^{in}\l^{jp}\l^{jq}\punkt\eqn
$$
The curvature term of course corresponds to a term with opposite sign in $\L$.

Surprising as it may seem,
it is possible to consistently introduce an arbitrary number of
fermions into the system
(\brackets) without destroying the supersymmetry. This is exactly what
happens when there are matter multiplets in the supersymmetric field theory.
The dynamics of the zero-modes of the matter fermions is completely
dictated by supersymmetry.
Consider real fermions
$\p^\a$ in a bundle over the riemannian manifold, transforming with some
connection $\w$, which in the context of fermion zero-modes will correspond
to the connection of the index bundle. With suitable choice of basis,
the bundle metric can be
chosen as $\d^{\a\b}$. We then add the brackets
$$
\eqalign{
&\{\p^\a,\p^\b\}=\d^{\a\b}\komma\cr
&\{V_m,\p^\a\}=\w_m^{\a\b}\p^\b\komma\cr}\eqn
$$
and add to the bracket $\{V_m,V_n\}$ a term $\half F_{mn\a\b}\p^\a\p^\b$,
with $F$ the field strength of $\w$. This is consistent with the Jacobi
identities. The $N\is\half$ supersymmetry
generator is still given by (\halfgenerator), and the hamiltonian
obtained as $\half\{Q,Q\}$ is
$$
\Ham=\half g^{mn}V_mV_n+{1\/4}F_{mn\a\b}\l^m\l^n\p^\a\p^\b\Eqn\hamwithF
$$
(in the case of vanishing $\w$, the $\p$'s are inert under supersymmetry,
which is consistent with their equations of motion $\dot\p\is0$).
The corresponding lagrangian is
$$
\L=\half g_{mn}\dot X^m\dot X^n+\half g_{mn}\l^mD_t\l^n+
	\half\p^\a D_t\p^\a-{1\/4}F_{mn\a\b}\l^m\l^n\p^\a\p^\b\komma
	\Eqn\QMwithF
$$
where $D_t\p^\a\is\dot\p^\a+\w^{\a\b}_m\dot X^m\p^\b$.
In the case of $N_f$ fundamental matter multiplets, the $\p$'s will
come in $2N_f$ copies of an $O(k)$ bundle over $\M_k$.
When target space is hyper-K\"ahler,
it is necessary for $F$ to be selfdual for (\QMwithF) to have extended
($N\is\half\cross4$) supersymmetry. Hitchin has
shown (see ref. [\MantonII]) that this is indeed the case for the
curvature of the index bundle of
zero-modes in the fundamental representation. The curvature of a
hyper-K\"ahler manifold is always selfdual (see eq\.\ (\finalR)).
Selfduality here means that
$F$ is a $(1,1)$-form with respect to any of the complex structures.
This definition agrees with the one used earlier when the riemannian
curvature of moduli space was calculated.

\section\moduliaction{The action for the moduli of N=2 SYM with matter}

In order to make a consistent truncation of the field-theoretic action
(\fieldtheoryaction), one approximates the motion in a $k$-monopole
background with slow motion in the bosonic and fermionic moduli (zero-modes),
a so called collective coordinate expansion [\Christ,\MantonI,\Rajaraman].
We expand the equations of motion in
$n\is \#({d\/dt})+\half\#(\hbox{fermions})$. At $n\is0$ one only has
the background fields $A$ with selfdual field-strength. At $n\is\half$,
there are the Weyl equations for the upper ($s$ chirality) components
of $\l$ and $\p$, which we denote $\a$ and $\b$, respectively.
Their lower ($c$ chirality) components vanish to this order.
The time dependence of the bosonic moduli is modeled so that
$A(x,t)=A(x,X(t))$. Then the equations at order $n\is1$ imply,
using $\dot A\is\dot X^m(\d_m A+D\e_m)$,
$$
\eqalign{
&A_0=\dot X^m\e_m+(D^*\!D)^{-1}(-\a^*\a+\half\b_\ff^*\cross\b_\f)\komma\cr
&A_5=(D^*\!D)^{-1}(\a^*\a+\half\b_\ff^*\cross\b_\f)\komma\cr
&q^*_\ff=-(D^*\!D)^{-1}(\a^*\b_\f)\punkt\cr}\Eqn\nisonesolution
$$
Inserting these solutions into the action (\fieldtheoryaction) and
discarding terms with $n\!>\!2$ gives the action (\QMwithF), after integrating
over three-space and using the expressions for connections and curvatures
on moduli space. There is also a constant topological term $-4\pi k$,
so the supersymmetry algebra contains a central charge at the classical level.
Since moduli space is hyper-K\"ahler, the action possesses
$N\is\half\cross4$ supersymmetry. We are amused to note that the second term in
(\indexcurvature) is present, and arises exactly from interaction with
the bosonic matter field $q$. In order to see this, one lets $s_m$ act on
the Weyl equation to obtain $\d_m A^*\r_\a=-D^*s_m\r_\a$ and uses the
expression for the higher mode projector (\projector) in the solution
(\nisonesolution) for $q$. Then, $Dq^*=\l^m\p^\a\Pi_+s_m\r_\a$.
A non-supersymmetric theory without matter
bosons would not get the correct curvature term for the fermionic moduli.
The result is of course valid also for $N\is4$
(which is $N\is2$ with one adjoint matter multiplet),
where earlier calculations [\Blum]
seem to have discarded the contribution from $q$ in the effective action
and approximated the curvature with its first term in (\curvature).
\vfill\eject

\section\conclusions{On the quantization of the effective action}

One would like to use the low energy effective action (\QMwithF) in order
to draw conclusions about the selfduality of the $N\is2$ theory with
four fundamental hypermultiplets. The details of this are left to a
forthcoming publication, but in order to put the present work in
some perspective, we would like to indicate the route to take.

The task is to find BPS-saturated states in the monopole spectrum.
The BPS limit for the mass is already present as the topological term
$4\pi k$ in the hamiltonian obtained from the field theory, so one
searches for zero energy states of the hamiltonian (\hamwithF).
The zero-modes of the adjoint fermions are divided into creation and
annihilation operators using the K\"ahler property of moduli space,
i.e\.\ by identifying one of the complex structures with $i$. Then,
in the case of pure $N\is2$,
the states may be identified with antiholomorphic forms in a Dolbeault
complex [\WittenI,\Alvarez,\GauntlettI,\SenI], and there is a nice
correspondence
with the cohomology of the manifold (which is severely restricted by the
hyper-K\"ahler property).

When the fundamental fermions and the field strength $F$ are present,
the situation changes slightly.
It is probably advantageous to treat the fundamental fermions with
gamma matrix quantization [\SeibergIII], analogous to the way the
fermions in the Ramond
sector of the superstring are treated.
Then the forms carry a $Spin(2kN_f)$ spinor index. The direct
identification of supercharges with exterior derivatives no longer holds,
since the fundamental fermions are affected by supersymmetry. Instead,
there will be an extra contribution behaving as a connection term.
We envisage that the zero energy states correspond to
forms which are harmonic with respect to the arising covariant laplacian.
The selfduality of $F$ should play an important r\^ole here.

In reference [\SeibergIII] part of the supposed mechanism behind the
expected selfduality of the model in question has been briefly sketched.
An $Sl(2,\Z)$ transformation on the coupling constant and $\theta$ angle
has to be accompanied by a $Spin(8)$ triality rotation (in the $k\is1$ sector)
to obtain a mapping between dual states.
One also has to remember that the masses induced by the Higgs mechanism
for the fundamental fields of the theory are different in the adjoint
and fundamental representation (since the BPS bound relates mass to charge).
This means that duality will map the different multiplets in
(\fieldtheoryaction) into sectors with different magnetic charge $k$.
We believe that a careful treatment
along the lines sketched here will verify these claims. This work is
in progress.

\vskip8\parskip
\underbar{Note added:}
Shortly after the completion of this work two papers
appeared [\Sehti,\GauntlettIII] that essentially depart from the effective
action derived in this paper and find the BPS-saturated spectrum for
magnetic charges $k\is1,2$ along the lines sketched in section 7. Their
results support the selfduality hypothesis for the $N\is2$ model with four
fundamental hypermultiplets.

\vfill\eject
\refout

\end